\documentstyle[preprint,aps]{revtex}
\tightenlines
\def\journal#1, #2, #3#4, #5#6#7#8    {
    {\rm #1~}{\bf #2}, #3#4 (#5#6#7#8)}

\def\prd{\journal Phys. Rev. D, }
\def\prl{\journal Phys. Rev. Lett., }

\def\npb{\journal Nucl. Phys. B, }
\def\plb{\journal Phys. Lett. B, }

\def\mpla{\journal Mod. Phys. Lett. A, }
\def\ijmpa{\journal Int. Jour. Mod. Phys. A, }
\def\jmp{\journal J. Math. Phys., }
\def\jpa{\journal J. Phys. A, }

\def\epj{\journal Eur. Phys. J., }

\newcommand{\beq}[1]{\begin{equation}\label{#1}}
\newcommand\eeq{\end{equation}}
\newcommand{\ba}[1]{\begin{eqnarray}\label{#1}}
\newcommand{\baa}{\begin{eqnarray}}
\newcommand\ea{\end{eqnarray}}
\newcommand{\bee}{\begin{equation}}
\def\nn{\nonumber \\}
\def\l{\lambda}
\def\n{\nu}

\def\o{\omega}

\def\tr{\tilde{\rho}}
\newcommand{\h}{Hamiltonian}

\def\hlf{\frac{1}{2}}

\newcommand{\pv}[1]{{-  \hspace {-4.0mm} #1}}
\newcommand{\col}{collective}

\begin{document}
\title{Calogero model, deformed oscillators and the collapse}
\author{Velimir Bardek$^1$, Larisa Jonke$^1$, Stjepan Meljanac$^1$, 
and Marijan Milekovi\'c$^2$
\footnote{e-mail: bardek@thphys.irb.hr, larisa@thphys.irb.hr,
meljanac@thphys.irb.hr, marijan@phy.hr} }
\address{$^1$Theoretical Physics Division,
Rudjer Bo\v skovi\'c Institute,\\ P.O. Box 180,
HR-10002 Zagreb, CROATIA\\
$^2$Theoretical Physics Department, Faculty of Science\\
P.O. Box 331, HR-10002 Zagreb, CROATIA}
\maketitle
\begin{abstract}
We discuss the behaviour of the Calogero model and the 
related model of $N$ deformed
oscillators with the $S_N$-extended Heisenberg algebra  for a special value of 
the constant of interaction/statistical parameter $\n$. The problem with finite 
number of deformed oscillators is analyzed in the algebraic approach, while 
collective-field theory is used to investigate the large-$N$ limit. 
In this limit, the  
system reduces to  a large number 
of collapsing (free) particles, for $\n=-1/N$.
\end{abstract}

PACS number(s): 03.65.Fd, 03.65.Sq, 05.30.Pr

The (rational) Calogero\cite{cal}
 model describes $N$ identical particles which interact
through an inverse-square two-body interaction and are subjected to a 
common confining harmonic force. The inverse-square potential can be regarded 
as a pure statistical interaction\cite{ha}
and the model maps to an ideal gas of particles obeying fractional statistics.
In one dimension, the usual notion of {\it exchange}  statistics is not well 
defined because the exchange necessarily involves  scattering\cite{lm}, 
but instead one can
apply  Haldane's definition of {\it exclusion} statistics\cite{haldane}.

The model is completely integrable in both the classical and quantum 
case\cite{per}, and that makes it an ideal "lab" for exploring new concepts in 
physics, like the already mentioned  {\it exclusion} statistics.
The discovery of the integrable spin-chain with the long-range interaction, the 
so-called Haldane-Shastry model\cite{hs} and its connection to the 
trigonometric 
version of the 
Calogero model, introduced the model to  condensed matter physics.
The Calogero model also appears when two-dimensional QCD   is 
formulated as a  random matrix model in the large-$N$ limit, restricted to 
a singlet subspace\cite{qcd}. More precisely, pure QCD in two dimensions
  is equivalent to the  
one-dimensional unitary-matrix model, 
which is a special case of the Calogero model.

Recently, the interest in the model 
has been renewed since it was proposed
that the
supersymmetric extension of the  model could provide a microscopic description
of the extremal Reissner-Nordstr\"om black hole\cite{bh}.
Many more applications of the model have been found, and
intensified the research of the model intrinsic properties.

In this letter we investigate the behaviour of the Calogero model and the 
related 
model of N deformed oscillators with the  $S_N$-extended Heisenberg algebra.
We apply the  Fock-space analysis and  collective-field 
theory. We discuss the physical picture of the problem with the negative 
statistical parameter $\nu$. To our knowledge, this is the first analysis of the
Calogero model and the related algebra of 
deformed oscillators for the negative, but still 
allowed $\nu$ leading to the positive definite Fock space, i.e. to 
quadratic-integrable wave functions, for general $N$. The two-particle case
was previously discussed in Refs.\cite{deV}.

The Calogero model is defined by the following \h :
\beq 1
H=-\frac{\hbar ^2}{2m}\sum_{i=1}^N\partial_i^2+\frac{m\o^2}{2}\sum_{i=1}^N x_i^2
+\frac{\hbar^2\n(\n-1)}{2m}
\sum_{i\neq j}^N\frac{1}{(x_i-x_j)^2},\eeq
where  $\n(\n-1)\geq -1/4$ 
is the dimensionless coupling constant and $\o$ is the 
strength of a harmonic confinement potential. In the following we
set $\hbar=m=\o=1$, for simplicity. 
The ground-state wave function is, up to normalization,
\beq 2
\psi_0(x_1,\ldots,x_N)=\theta(x_1,\ldots,x_N)\exp\left(-\frac{1}{2}
\sum_{i=1}^N x_i^2\right),\eeq
where
$\theta(x_1,\ldots,x_N)=\prod_{i<j}^N(x_i-x_j)^{\n}$,
with the corresponding ground-state energy $E_0=N[1+(N-1)\n]/2$.
The physical requirement of vanishing of the wave function $\psi_0(x_1,\ldots,
x_N)$ for coincidence points implies $\nu >0$.
We note in passing that for $0<\nu\leq 1/2$ one needs to renormalize the theory,
see Refs.\cite{reno}.

One can introduce the following analogs of creation and annihilation
operators\cite{all}:
\beq 4
a_i^{\dagger}=\frac{1}{\sqrt{2}}(-D_i+x_i),\;
a_i=\frac{1}{\sqrt{2}}(D_i+x_i), \eeq
where
$$ D_i=\partial_i+\n\sum_{j,j\neq i}^N\frac{1}{x_i-x_j}(1-K_{ij})$$
are Dunkl derivatives.
The elementary generators $K_{ij}$ of the symmetry group $S_N$ exchange
labels $i$ and $j$:
\ba 5
&&K_{ij}x_j=x_iK_{ij},\; K_{ij}=K_{ji},\; (K_{ij})^2=1,\nn
&& K_{ij}K_{jl}=K_{jl}K_{il}=K_{il}K_{ij},\;{\rm for}\;i\neq j,\;
i\neq l,\;j\neq l,\nonumber\ea
and we choose $K_{ij}|0\rangle=|0\rangle$.
One can easily check that the commutators of the creation and
annihilation operators (\ref{4}) are
\beq 6
[a_i,a_j]=[a_i^{\dagger},a_j^{\dagger}]=0,\;
[a_i,a_j^{\dagger}]=\left(1+\n\sum_{k=1}^NK_{ik}\right)\delta_{ij}
-\n K_{ij}\equiv A_{ij}.
\eeq
The algebra (\ref{6}) is consistent for all values of the parameter $\nu$.
The only restriction on the parameter $\nu$ comes from the representation
of the algebra on the Fock space.
By performing a similarity transformation of the \h \ (\ref{1}), we
obtain the reduced \h \
\beq 7
H'=\theta^{-1}H\theta=\hlf\sum_{i=1}^N\{a_i,a_i^{\dagger}\}=\sum_{i=1}^N
a_i^{\dagger}a_i+E_0,\eeq when restricted to symmetric functions and
acting as a total number operator.
The physical space of $N$ identical bosons is a space of totally symmetric
functions
(for $N$ identical fermions one  chooses
a space of totally antisymmetric functions).
We can then rewrite the \h \ $H'$  as a \h \ for deformed 
oscillators\cite{wess}:
\bee\label{defor}
H'=\frac{1}{2}\sum_{i=1}^N(p_i^2+x_i^2),\eeq
where $p_i=-i D_i=-i(a_i-a_i^{\dagger})/\sqrt{2}$ is the hermitian momentum 
operator,  $x_i=(a_i+a_i^{\dagger})/\sqrt{2}$ and the deformed commutation
 relation is  
$$[x_i,p_j]=i\left(1+\n\sum_{k=1}^NK_{ik}\right)\delta_{ij}
-i\n K_{ij}.$$

Let us now investigate the condition under which the states  in the  complete
Fock space of algebra (\ref{6}) have positive norm. 
To this end, we calculate the general matrix element of the Gram  
matrix\cite{jos}
for one-particle excited states:
\beq 8
\langle 0|a_ia_j^{\dagger}|0\rangle =\langle 0|A_{ij}|0\rangle =
\delta_{ij}(1+N\n)-\n,\eeq
where we have used the fact that $a_i|0\rangle =0$ and $\langle 0|0\rangle=1$. 
There is only one non-degenerate eigenvalue
$\l_1=1/N$ with the corresponding  eigenvector $e_1=\sum_{i=1}^Na_i^{\dagger}
|0\rangle$, and $N-1$ degenerate eigenvalues $\l_{N-1}=1+N\n$ with
 eigenvectors $e_{N-1}=(a_1^{\dagger}-a_i^{\dagger} )|0\rangle,\;
i=2,3,\ldots,N$. The Gram matrix is positive definite, i. e. there are 
no vectors with negative norm if all eigenvalues are positive. 
Thus
$$1+N\n>0\;\Rightarrow\;\n>-\frac{1}{N}.$$
At $\n=-1/N$ there is a critical point where the states $(a_i^{\dagger}
-a_j^{\dagger} )|0\rangle$ have null norm\footnote{If we had extracted the 
prefactor $\theta(x_1,\ldots,x_N)=\prod_{i<j}^N(x_i-x_j)^{1-\n}$ 
from the wave function, we would have obtained the critical 
point $\nu=1+1/N$.}. 
It should be mentioned that the necessary condition for positivity
is $\langle 0|a_ia_i^{\dagger}|0\rangle>0 $, i. e. $\n>-1/(N-1)$.
However, this condition is not sufficient. Namely, 
for $-1/(N-1)<\n<-1/N$, the states $(a_i^{\dagger}-a_j^{\dagger} )|0\rangle$ 
have negative norm, although the mean value $\langle 0|a_i
a_i^{\dagger}|0\rangle $ is positive. One can show that for two-, three- 
and many-particle states, the same condition for positivity of 
eigenvalues, i. e.  $\n>-1/N$, is required\cite{mms}. 
There exists the universal critical point at $\n=-1/N$, where all matrix 
elements of the arbitrary k-multi-state matrix are equal to $k!/N^k$. Namely, 
for one-particle states, $\langle 0|a_ia_j^{\dagger}|0\rangle=1/N$
at the critical point $\n=-1/N$.
For two-particle excited states, we obtain 
\bee\label{2p}
\langle 0|a_{i_2}a_{i_1}a_{j_1}^{\dagger}a_{j_2}^{\dagger}|0\rangle=
\langle 0|a_{i_2}A_{i_1,j_1}a_{j_2}^{\dagger}|0\rangle+
\langle 0|a_{i_2}a_{j_1}^{\dagger}A_{i_1,j_2}|0\rangle=\frac{2}{N^2},\eeq
since  $\langle 0|A_{i,j}|0\rangle=1/N$ and
$\langle 0|a_{i_2}A_{i_1,j_1}a_{j_2}^{\dagger}|0\rangle=1/N^2$ at the 
critical point. 
It can be shown by induction that the general matrix element at $\n=-1/N$ is 
given by
\bee\label{genm}
\langle 0|a_{i_k}\cdots a_{i_1}a_{j_1}^{\dagger}\cdots 
a_{j_k}^{\dagger}|0\rangle=
\frac{k!}{N^k},\;\forall k,\;N\geq 2.\eeq
If $\n>-1/N$, the diagonal matrix elements satisfy the inequality
$$\langle 0|a_{i_k}\cdots a_{i_1}a_{i_1}^{\dagger}\cdots
a_{i_k}^{\dagger}|0\rangle>\frac{k!}{N^k},$$
which ensures the positivity condition.

What is the interpretation of these results like? For $\n>-1/N$, we see that 
all k-states have positive norm. The corresponding Gram matrix of order $N^k$ 
shows $N$ independent $\n$-deformed oscillators which can be mapped to 
$N$ bosons and vice versa\cite{mms}. However, at the critical point there is 
only one state for any $k$ with the eigenvalue $k!>0$. 
All other eigenvalues vanish!
This means that in the limit $\n N\rightarrow -1$ 
the system of deformed oscillators exhibits
singular behaviour. There survives only one oscillator $\sum_{i=1}^N
a_i^{\dagger}/\sqrt{N}$ describing the motion of the centre-of-mass 
coordinate. All other $N-1$ relative coordinates collapse into the same 
point - the centre of mass. One can easily check that $\tilde a_i^{\dagger}
|0>=0$ and therefore $\tilde p_i|0>=\tilde x_i|0>=0$, where "$\sim $" denotes
operators defined in the centre-of-mass system. For example, 
$\tilde a_i^{\dagger}=a_i^{\dagger}-\sum_ja_j^{\dagger}/N$. This means that 
the relative coordinates and the
 relative momenta are zero at the critical point. 
Also, the relative energy vanishes for this critical value of the 
parameter $\nu$.
This collapse in the deformed quantum-mechanical system
 can be regarded as a toy model for confinement.

The interval $(-1/N,0)$ for the 
statistical parameter $\nu$ is physically acceptable
for the \h \ $H'$ describing $N$ oscillators with the 
$S_N$-extended Heisenberg
algebra, but it is not allowed for the original Calogero \h .
Hence, there is no collapse of Calogero particles.
The wave function of the ground state (\ref{2}) contains the Jastrow 
factor, i. e. the product of all relative coordinates with $\nu$ as a power.
Consequently, the wave functions diverge for the negative value of $\nu$, at 
coincidence points, although  for $\nu>-1/N$ they remain normalizable.
The \h \ $H'$ obtained by similarity transformation (\ref{7}) leads to the 
deformed oscillators with $S_N$-extended Heisenberg algebra and is 
equivalent to the Calogero model only for $\nu>0$.

Starting from the symmetric Fock space and the 
algebra of $S_N$-symmetric observables in the Calogero model,
one can construct the mapping to free Bose oscillators (and vice versa) for 
$\n>-1/N$\cite{PLB}.
Then it immediately follows that the partition function is 
$${\cal Z}=e^{\beta E_0}\prod_{k=1}^N\frac{1}{1-e^{-\beta k}}.$$
However, all observables disappear at the collapse point, leaving only the  
centre-of-mass coordinate, so the partition function reduces to the
partition function of a single oscillator ${\cal Z}_{\rm crit}=\exp{(\beta/2)}
/(1-\exp{(-\beta)})$.

In order to check the consistency and further elucidate our 
physical picture of the behaviour of the system at the critical point, we 
now turn to the case of infinite number of particles.
Let us consider the large$-N$ Calogero model in the \h \ collective-field 
approach\cite{coll} 
based on the $1/N$ expansion. It has been shown, that in the 
large$-N$ limit, the \h \  $H'$ can be expressed entirely in terms of density 
of particles $\rho(x)$ and its canonical conjugate $\pi(x)=-i\delta/\delta\rho
(x)$. The Jacobian $J$ of the transformation from $x_i$ into $\rho(x)$ 
rescales the symmetric wave functional
$$\phi(x_1,x_2,\ldots,x_N)=\sqrt{J}\phi(\rho)$$ 
resulting in the hermitian collective-field \h \
\bee \label{h3}
H=\frac{1}{2}\int dx\rho(x)(\partial_x\pi)^2+
\frac{1}{8}\int dx\rho(x)\left(\partial_x\frac{\delta\ln J}{\delta\rho(x)}
\right)^2
-\frac{1}{2}\int dx\rho(x)x^2, \eeq
and the Jacobian determined from the hermicity condition
\bee \label{hy}
\partial_x\left(\rho(x)\partial_x\frac{\delta\ln J}{\delta\rho(x)}\right) =
(\n-1)\partial_x^2\rho(x)+2\n\partial_x\left(\rho(x)\pv{\int} dy
\frac{\rho(y)}{x-y}\right) .\eeq
The first term in the collective \h \ (\ref{h3}), quadratic in the conjugate 
momentum $\pi(x)$, represents the kinetic energy of the problem. 
To find the ground-state energy of our system, we assume that the corresponding
 \col-field configuration is static and has a vanishing momentum $\pi$.
Therefore, the leading part of the \col-field \h\ in the $1/N$ expansion is
given by the effective potential 
\bee \label{v1} 
V_{\rm eff}(\rho)=
\frac{1}{8}\int dx\rho(x)\left[(\n-1)\frac{\partial_x\rho(x)}
{\rho(x)}+2\n\pv{\int} dy\frac{\rho(y)}{x-y}\right]^2 +\frac{1}{2}
\int dx\rho(x)x^2.\eeq
This can be rewritten as a complete square, up to the 
numerical constant (the ground-state energy):
\bee \label{v2} 
V_{\rm eff}(\rho)=
\frac{1}{8}\int dx\rho(x)\left[(\n-1)\frac{\partial_x\rho(x)}
{\rho(x)}+2\n\pv{\int} dy\frac{\rho(y)}{x-y}-2x\right]^2 +\frac{N}{2}
[\n(N-1)+1].\eeq
The first term in (\ref{v2}) is positive semidefinite,
and, therefore, its contribution to
the ground-state energy vanishes if there exists a positive solution of the
first-order differential Bogomol'nyi-type equation:
\bee \label{b1}
(\n-1)\partial_x\rho(x)+2\n\rho(x)\pv{\int} dy\frac{\rho(y)}{x-y}-2x
\rho(x)=0.\eeq
What happens at the critical value $\n=-1/N$? 
Since the collective field obeys
the normalization condition $\int dx\rho(x)=N$, 
it is legitimate to rescale $\rho(x)=N\tr(x)$. Rewriting Eq.(\ref{b1})
for the rescaled collective field $\tr(x)$, we obtain 
at the critical point, in the leading approximation in $1/N$:
\bee\label{b2}
\partial_x\tr(x)+2\tr(x)\pv{\int} dy\frac{\tr(y)}{x-y}+2x\tr(x)=0.\eeq
Having in mind the identity for the singular distribution\cite{sch}:
\bee\label{sing}
\frac{\partial \delta(x)}{\partial x}+2\delta(x)P\frac{1}{x}=0,\eeq
where P denotes the principal distribution, it is evident that the
normalizable solution of Eq.(\ref{b2}) is given by the $\delta$-function,
$\tr(x)=\delta(x)$.
All particles are concentrated around the origin with density given by 
\bee\label{de}
\rho(x)=N\delta(x).\eeq
The corresponding ground-state energy reduces to 
$E_0=1/2$, i. e. to the ground state of the single harmonic oscillator. 

Let us briefly summarize the main findings of the collective-field approach.
We confirm the existence of the critical point $\n=-1/N$ found using
 the exact
algebraic
approach valid for any $N$. Also, we observe the collapse of the
system accompanied by the simultaneous emerging of a single free oscillator.

We point out that for $N\rightarrow\infty$, $\nu=-1/N$, the Fock space 
representation of the $S_N$-extended Heisenberg algebra (\ref{6}) reduces 
to the Fock space of the centre-of-mass oscillator, although the 
$S_N$-extended Heisenberg algebra becomes the ordinary Heisenberg 
 algebra for infinitely many oscillators. 
Hence, for $\nu=-1/N$, the 
 system reduces to a large number of collapsing (free)
particles.
On the other hand, the 
limit $\nu\rightarrow 0_{+}$ is  regular, i.e. the Fock space corresponds to 
the ordinary bosons for the ordinary Heisenberg algebra.

The behaviuor of the system near the critical point 
 is a universal feature of a large class of 
permutation-invariant algebras\cite{mms}, which can be written as
$$a_ia_j^{\dagger}-qa_j^{\dagger}a_i=\delta_{ij}[1+\sum_{l=1}^N\n_{il}K_{il}]-
\n_{ij}K_{ij},\;|q|\leq 1.$$ 
When all $\n_{ij}$ tend to the critical point $-1/N$, only a single 
q-deformed oscillator,  representing the centre of mass, survives. 
It gives us a new and interesting form of matter.
We hope that these ideas
might be applied to problems like black-hole physics, extremely 
dense matter, anyons and confinement.  
 
Acknowledgment

We thank the referee for critical consideration of the manuscript.
This work was supported by the Ministry of Science and Technology of the
Republic of Croatia under contracts No. 00980103 and 0119222.

\end{document}